\documentclass[aps,prb,reprint,amssymb,amsmath,superscriptaddress]{revtex4-1}

\usepackage{graphicx}
\usepackage{epstopdf}
\usepackage[english]{babel}

\begin{document}

\title{Weak antilocalization of holes in HgTe quantum wells with a normal energy spectrum}

\author{G.~M.~Minkov}
\affiliation{M.~N.~Miheev Institute of Metal Physics of Ural Branch of
Russian Academy of Sciences, 620137 Ekaterinburg, Russia}

\affiliation{Institute of Natural Sciences, Ural Federal University,
620002 Ekaterinburg, Russia}

\author{A.~V.~Germanenko}

\author{O.~E.~Rut}
\affiliation{Institute of Natural Sciences, Ural Federal University,
620002 Ekaterinburg, Russia}

\author{A.~A.~Sherstobitov}
\affiliation{M.~N.~Miheev Institute of Metal Physics of Ural Branch of
Russian Academy of Sciences, 620137 Ekaterinburg, Russia}

\affiliation{Institute of Natural Sciences, Ural Federal University,
620002 Ekaterinburg, Russia}

\author{S.~A.~Dvoretski}

\affiliation{Institute of Semiconductor Physics RAS, 630090
Novosibirsk, Russia}

\author{N.~N.~Mikhailov}

\affiliation{Institute of Semiconductor Physics RAS, 630090
Novosibirsk, Russia}

\date{\today}

\date{\today}

\begin{abstract}
The results of experimental study of interference induced
magnetoconductivity in narrow HgTe quantum wells  of hole-type
conductivity with a normal energy spectrum are presented.
Interpretation of the data is performed with taking into account the
strong spin-orbit splitting of the energy spectrum of the
two-dimensional  hole subband. It is shown that the phase relaxation
time  found from the analysis of the shape of magnetoconductivity
curves for the relatively low conductivity when the Fermi level lies in
the monotonic part of the energy spectrum of the valence band behaves
itself analogously to that observed in narrow HgTe quantum wells of
electron-type conductivity. It increases in magnitude with the
increasing conductivity and decreasing temperature following the $1/T$
law. Such a behavior corresponds to the inelasticity of
electron-electron interaction as the main mechanism of the phase
relaxation and agrees well with the theoretical predictions. For the
higher conductivity, despite the fact that the dephasing time remains
inversely proportional to the temperature, it strongly decreases with
the increasing conductivity. It is presumed that a nonmonotonic
character of the hole energy spectrum could be the reason for such a
peculiarity. An additional channel of the inelastic interaction between
the carriers in the main and secondary maxima occurs when the Fermi
level arrives the secondary maxima in the depth of the valence.
\end{abstract}


\maketitle

\section{Introduction}
\label{sec:intr} New type of two-dimensional (2D) systems, which energy
spectrum is formed by the spin-orbit interaction has attracted
considerable interest during the last decade.  The structures with HgTe
quantum well (QW) hold special place among such structures. The strong
intrinsic spin-orbit interaction leads to the energetic inversion of
the $\Gamma_8$ and $\Gamma_6$ bands in the bulk spectrum of HgTe. The
$\Gamma_6$ band, which is the conduction band in the usual
semiconductors, is located lower in energy than the $\Gamma_8$ band so
that the last forms both the conduction and valence bands, which are
not separated by the gap. It results in nontrivial dependence of the
energy gap on the width of quantum well ($d$) and causes other
important features of the energy spectrum. So the energy spectrum in
CdTe/HgTe/CdTe quantum well  at $d=d_c\simeq 6.5$~nm is
gapless\cite{Gerchikov90} and is close to the linear Dirac-like
spectrum  at small quasimomentum ($k$).\cite{Bernevig06} When the HgTe
layer is thin, $d<d_c$, the ordering of energy subbands of spatial
quantization is analogous to that in conventional narrow-gap
semiconductors; the highest valence subband at $k=0$ is formed from the
heavy hole $\Gamma_8$ states, while the lowest electron subband is
formed both from the $\Gamma_6$ states and light hole $\Gamma_8$
states. For thick HgTe layer, $d>d_c$, the quantum well is in the
inverted regime; the main electron subband is formed from the heavy
hole $\Gamma_8$ states,\cite{Dyak82e} whereas the subband formed from
the $\Gamma_6$ states and light hole $\Gamma_8$ states sinks into the
valence band. This in turn leads to a significant modification of the
kinetic phenomena and to arising of new ones. Besides, the energy
spectrum of HgTe based heterostructures is very sensitive to a
structure asymmetry due to strong Bychkov-Rasba effect.\cite{Rash84}

The effects that depend not only on the energy spectrum, but on the
wave functions also, have even more strong peculiarities. The
interference contribution to the conductivity is just this effect.  The
suppression of interference by the magnetic field leads to the arising
of the low-magnetic-field weak-localization (WL) or
weak-antilocalization (WaL) magnetoconductivity (MC).	Experimentally,
the low field magnetoconductivity of 2D electron gas in HgTe QW's was
observed in Refs.~\onlinecite{Olshanetsky10,Muhlbauer14} and
investigated in detail in Refs.~\onlinecite{Minkov12,Minkov13-1}. There
was shown\cite{Minkov13-1} that the MC curves for the structures with
normal spectrum, $d<d_c$, are well described by the conventional
Hikami-Larkin-Nagaoka (HLN) expression\cite{Hik80} within a wide
conductivity range. The phase relaxation time ($\tau_\phi$) found from
the fit of MC curves to this expression increases with the temperature
decrease as $1/T$, that corresponds to the case when inelasticity of
electron-electron (\emph{e-e}) interaction is the main dephasing
mechanism.\cite{AA85} It is important that the $\tau_{\phi}$ value
increases with increasing conductivity and this dependence is close to
that predicted for this dephasing mechanism:\cite{Nar02}
$\tau_\phi\propto\sigma/\ln{\sigma}$, where $\sigma$ is the
conductivity measured in units of $G_0=e^2/(2\pi^2 \hbar)$. Thus, the
interference  correction to the conductivity of electron 2D gas in HgTe
QW's  with $d<d_c$ behaves the same as the correction in the usual 2D
structures.

Another behavior of $\tau_\phi$  was observed for  electrons in the
structures with inverted spectrum,  $d=(9-10)$~nm.\cite{Minkov12}
Whereas the temperature dependence of $\tau_\phi$ remains conventional,
$\tau_\phi\propto 1/T$, the  $\sigma$ dependence of $\tau_\phi$ is
strange: $\tau_\phi$ is practically independent of conductivity.  And
this occurs in spite of the fact that MC is well described by the HLN
expression also. The reason of such a phenomenon is yet to be
explained.

Concerning the weak localization in the hole 2D HgTe based systems, the
theories\cite{Tkachov11,Gornyi14} predict that the interference quantum
correction  for electrons and holes should be close to each other both
for $d\lesssim d_c$ and $d\gtrsim d_c$. This is because the energy
spectra of the conduction and valence bands near critical point $d=d_c$
are identical for $d\lesssim d_c$ and $d\gtrsim d_c$ for small
quasimomentum values.\cite{Bernevig06} Experimental studies of the weak
localization in the hole HgTe QW's are absent to the best of our
knowledge.

In this paper we present the results of experimental investigation of
interference induced  magnetoconductivity in the gated HgTe quantum
wells of hole-type conductivity with normal energy spectrum. The
measurements were performed on the same heterostructures, whose energy
spectra were studied in Ref.~\onlinecite{Minkov14}. The specific
feature of the samples  is the strong spin-orbit splitting of the
spectrum due to the asymmetry of the quantum well. Analyzing the shape
of the magnetoconductivity curves with taking this fact into account we
obtain the phase relaxation time within a wide conductivity range at
different temperatures. We show that the phase relaxation time
increases in magnitude with the decreasing temperature following the
$1/T$ law indicating that the inelasticity of \emph{e-e} interaction is
the main mechanism of the phase relaxation. The conductivity dependence
of $\tau_\phi$ is also usual for such a dephasing mechanism at low
conductivity, $\sigma\lesssim 100\,G_0$; the dephasing time increases
with the increasing conductivity. For the higher conductivity, the
behavior of $\tau_\phi$ changes drastically. It strongly decreases with
$\sigma$. The experimental results are discussed having regard to
nonmonotonic character of the hole dispersion law.

\section{Experimental}
\label{sec:expdet}

Our samples with HgTe quantum wells  were realized on the basis of two
HgTe/Hg$_{1-x}$Cd$_{x}$Te ($x=0.55-0.65$) heterostructures grown by
molecular beam epitaxy on GaAs substrate with the (013) surface
orientation.\cite{Mikhailov06} The nominal width of the quantum well
was   $5.8$~nm and  $5.6$~nm  in the structures H724 and H1122,
respectively. The quantum wells  were of hole-type  conductivity. The
results for these structures are similar and we will mainly discuss the
results which were obtained for the structure H724 with the higher Hall
mobility. The samples were mesa etched into standard Hall bars of
$0.5$~mm  width and the distance between the potential probes of
$0.5$~mm. To change and control the hole density ($p$) in the quantum
well, the field-effect transistors were fabricated with parylene as an
insulator and aluminium as a gate electrode. For each heterostructure,
four samples were fabricated and studied. The hole density was about
$1\times 10^{11}$~cm$^{-2}$ for zeroth gate voltage and something less
in the structure H1122. The measurements were performed at temperature
of $1.3-4.2$~K.

\section{Results and discussion}
\label{sec:res}

The WL correction and MC curves depend not only on the momentum-,
phase-, and spin-relaxation times but on the energy spectrum also.
Therefore, before discussing the low-field magnetoconductivity let us
look the hole energy spectrum of the structures under study that was
restored when analyzing the data of the transport measurements in
Ref.~\onlinecite{Minkov14}.

\begin{figure}
\includegraphics[width=\linewidth,clip=true]{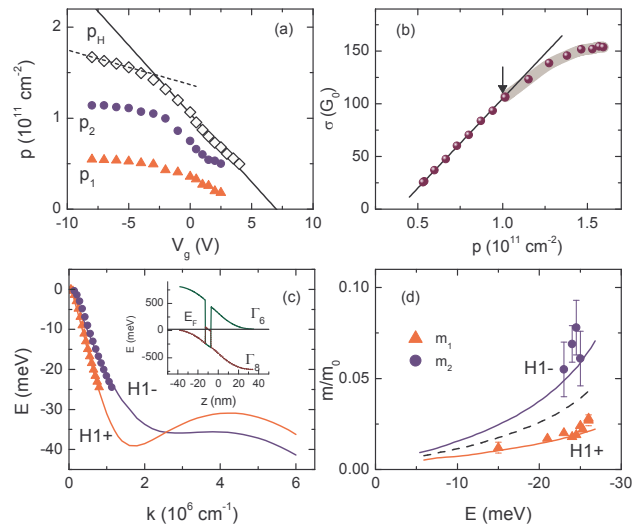}
\caption{(Color online) (a) -- The gate voltage dependence of the Hall
density $p_\text{H}=1/[eR_\text{H}(0.1\text{ T})]$ (diamonds) and densities $p_1$ and $p_2$
(triangles and circles, respectively) in the spin split subbands found from the SdH oscillations (for more
details, see Ref.~\onlinecite{Minkov14}). The solid
line is drawn with the slope $-1.5\times 10^{10}$~cm$^{-2}$V$^{-1}$, the dotted line is provided as a guide to the eye.
(b) -- The conductivity plotted against the hole density. The data shadowed correspond
to the regime where $\tau_\phi$ drops with the increasing conductivity [see. Fig.~\ref{f5}(a)].
(c) -- The energy spectrum of the valence band.
Symbols are restored from the experimental data.\cite{Minkov14} The lines are the results of theoretical
calculation with taking into account the electric field in the well. The inset shows the energy diagram of the structure calculated
under the assumption that acceptor and donor densities in the lower and
upper barriers, are $3\times 10^{17}$~cm$^{-3}$. (d) --  The energy
dependencies of the effective masses for the H1+ and H1$-$ hole subbands.
The symbols are the data,\cite{Minkov14}
the solid lines are the calculation results. The dashed curve is the average value
of the effective mass $m_\text{av}=(m_1+m_2)/2$.}\label{f1}
\end{figure}

The gate voltage dependence of the Hall density,
$p_\text{H}=1/[eR_\text{H}(0.1\text{~T})]$, where $R_\text{H}$ is the
Hall coefficient $R_\text{H}=\rho_{xy}/B$,  and the hole densities that
were found from the periods of Shubnikov-de Haas (SdH) oscillations are
shown in Fig.~\ref{f1}(a). One can see that $p_\text{H}$ linearly
changes with $V_g$ with a slope $dp_\text{H}/ dV_g$ of about
$-1.5\times 10^{10}$~cm$^{-2}$V$^{-1}$ at $-2.5<V_g<4$~V, where the
hole density is less than $\simeq1.5\times 10^{11}$cm$^{-2}$. At
$V_g\lesssim -3.5$~V, the slope becomes much less. Note that
capacitance between the gate electrode and the 2D channel measured on
the same sample is constant over the whole gate voltage range so that
the value of $C/e=(1.4\pm 0.15)\times 10^{10}$cm$^{-2}$ is very close
to $|dp_\text{H}/ dV_g|=1.5\times10^{10}$~cm$^{-2}$V$^{-1}$. Analyzing
these data together with the data obtained from the analysis of the
temperature dependences of the SdH oscillations amplitude, we have
reconstructed the energy spectrum near the valence band top in
Ref.~\onlinecite{Minkov14}. These results are reproduced in
Fig.~\ref{f1}(c). One can see first of all that the valence band is
strongly split by spin-orbit interaction, so that the ratio of the hole
densities in the subbands is approximately equal to two. The energy
spectrum is strongly non-parabolic, i.e., the hole effective masses
significantly increases with the energy increase [Fig.~\ref{f1}(d)].
These results are well described within the framework of the \emph{kP}
model if one supposes that the lower barrier remains of \emph{p} type,
while the upper one is converted to the \emph{n} type after the growth
stop, so that the quantum well is located in a strong electric field of
\emph{p-n} junction. The other key feature of the calculated spectrum
is the secondary maxima located at $k\simeq 4\times 10^6$~cm$^{-2}$ at
an energy distance of about $30$~meV from the main maxima. As suggested
in Ref.~\onlinecite{Minkov14} these maxima can be responsible for
flattening of the $p_\text{H}$~vs~$V_g$ dependence at $V_g\lesssim
-3.5$~V.

\begin{figure}
\includegraphics[width=1.0\linewidth,clip=true]{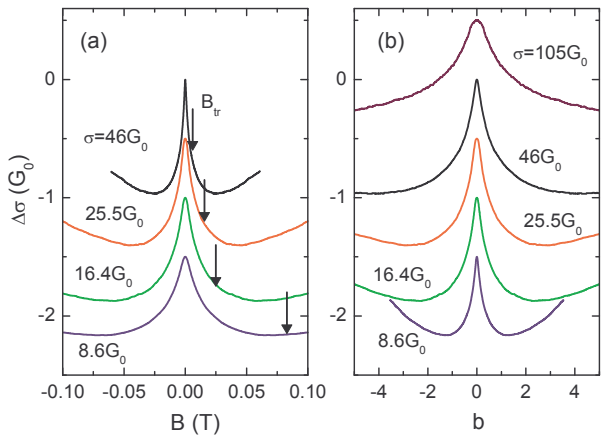}
\caption{(Color online) (a) -- The magnetic field dependences of $\Delta\sigma$  for
different conductivity values at $T=1.35$~K.
The arrows show the values of $B_{tr}$. (b) -- The same
data plotted against  the relative
magnetic field $b=B/B_{tr}$. For clarity, the curves are shifted
in the vertical direction.
}\label{f2}
\end{figure}

Now we are in position to analyze the low-field magnetoconductivity
$\Delta\sigma(B)=1/\rho_{xx}(B)- 1/\rho_{xx}(0)$. These dependencies
for different conductivities controlled by the applied gate voltage are
shown in Fig.~\ref{f2}. It is seen that the negative
magnetoconductivity (antilocalization behavior) is observed at low
magnetic field. In the higher magnetic field, MC reverses the sign
demonstrating the localization behavior. As evident the higher is the
conductivity, the lower is this field at which such a crossover occurs.

It is well known that the characteristic field for the weak
localization is a transport magnetic field $B_{tr}=\hbar/(2e l^2)=\pi^3
\hbar p/e\times (G_0/\sigma)^2$ with $l$ as the transport mean free
path, therefore in Fig.~\ref{f2}(b) we have plotted $\Delta\sigma$
against $b=B/B_{tr}$.\footnote{See Section~A in Supplemental Material
for justification of using of the one-type-carrier approximation to
analysis of weak antilocalization magnetoconductivity in the case of
strong spin-orbit splitting} As evident the crossover from the
antilocalization to localization behavior of magnetoconductivity takes
place at $b\gtrsim 1$  for all the conductivity values. Since the
minimum is observed in relatively high magnetic field $B\simeq
(0.02-0.07)$~T [Fig.~\ref{f2}(a)], it is quite possible that a
different mechanism, for example, the \emph{e-e} interaction correction
to the conductivity is responsible for the change of the MC sign.

The quantitative analysis of the magnetic field dependence of the
conductivity resulting from the suppression of the interference
correction by the magnetic field is not a simple problem for our case
because the valence band is strongly split by  spin-orbit interaction.
Since the hole effective masses in spin subbands are different, the
mobilities can be different and, hence, the transport magnetic fields
as well as the $\tau$ to $\tau_\phi$ ratio can be different also. To
the best of our knowledge the WL magnetoconductivity for such a
specific situation is as yet not calculated. However, there are two
limiting cases when the expression for the magnetoconductivity can be
obtained from qualitative considerations: (i) the transitions between
subbands are slow as compared with the dephasing processes,
$\tau_{12}\gg \tau_\phi$, where $\tau_{12}$ is the transition time, and
(ii) they are fast, $\tau_{12}\ll \tau_\phi$.

In the first case, $\tau_{12}\gg \tau_\phi$,  the total WL
magnetoconductivity is simply the sum of independent contributions from
each spin subbands:
\begin{equation}
 \Delta\sigma = \Delta\sigma_1+\Delta\sigma_2.
 \label{eq05}
\end{equation}
Here, $\Delta\sigma_1$ and $\Delta\sigma_2$ are antilocalizing because
the carrier spin is rigidly coupled with the momentum due to the
Bychkov-Rashba effect so that:\cite{Gornyi98}
\begin{equation}
 \Delta\sigma = -\frac{G_0}{2}\left[{\cal H}\left(\frac{B}{B_{tr}^{(1)}},\frac{\tau_1}{\tau_\phi^{(1)}}\right)
+  {\cal H}\left(\frac{B}{B_{tr}^{(2)}},\frac{\tau_2}{\tau_\phi^{(2)}}\right)\right],
 \label{eq10}
\end{equation}
where
\begin{eqnarray}
{\cal H}(b,x) &= &\psi\left(\frac{1}{2}+
 \frac{x}{b}\right)-\psi\left(\frac{1}{2}+
 \frac{1}{b}\right)-\ln{x} \nonumber \\
 &\simeq &\psi\left(\frac{1}{2}+
 \frac{x}{b}\right)-\ln{\left(\frac{x}{b}\right)},\,\,\, b\lesssim 1, \nonumber
\end{eqnarray}
$\psi(x)$ is a digamma function, and $i=1,2$ numbers the spin subbands.
Obviously it is impossible to find reliably the dephasing times while
fitting the smooth curve, if the subband parameters entering
Eq.~(\ref{eq10}) is strongly different and unknown independently with
high enough accuracy. It becomes possible only when the corresponding
parameters of subbands are close to each other. Then, in the first
approximation, Eq.~(\ref{eq10}) is reduced to
\begin{equation}
\Delta\sigma = \alpha G_0 {\cal H}\left( \frac{B}{B_{tr}}, \frac{\tau}{\tau_\phi}\right)
 \label{eq30}
\end{equation}
with $\alpha\simeq -1$, which can be already used to find $\tau_\phi$.

If a carrier executes many transitions between subbands within the
phase breaking time, i.e., $\tau_{12}\ll \tau_\phi$, the
magnetoconductivity has the same form, Eq.~(\ref{eq30}), in which,
however, the prefactor $\alpha$ is equal to $-1/2$ instead of $-1$, and
the parameters $B_{tr}$, $\tau$, and $\tau_\phi$ are averages over two
subbands.\cite{Averkiev98,Raichev00} Qualitatively, this can be
explained by the fact that the probability of return to the starting
point after traveling over closed path, while remaining in the same
subband, is reduced by half.

Our analysis of the magnetic field dependences of the resistivity
components  $\rho_{xx}$ and $\rho_{xy}$ performed in
Ref.~\onlinecite{Note1} within classical magnetic field range  allows
us to estimate the hole densities and mobilities in the different spin
subbands and, thus, estimate the $\tau_i$ and $B_{tr}^{(i)}$ values. It
turns out that the values of $\tau_1 B_{tr}^{(1)}$ and $\tau_2
B_{tr}^{(2)}$, which are determined the run of MC curve [see
Eq.~(\ref{eq10})] at $b<1$, are close to each other within accuracy
better than $30$~\% over the whole conductivity range. Besides,
simulating the WaL MC curves we show \emph{ibidem} that the use of
one-band formula, Eq.~(\ref{eq30}), allows us to obtain the dephasing
rate with accuracy better than $10$~\% both for $\tau_{12}\ll
\tau_\phi$ and $\tau_{12}\gg \tau_\phi$. Thus, the use of
Eq.~(\ref{eq30}) for the data analysis seems warranted in our case.

\begin{figure}
\includegraphics[width=1.0\linewidth,clip=true]{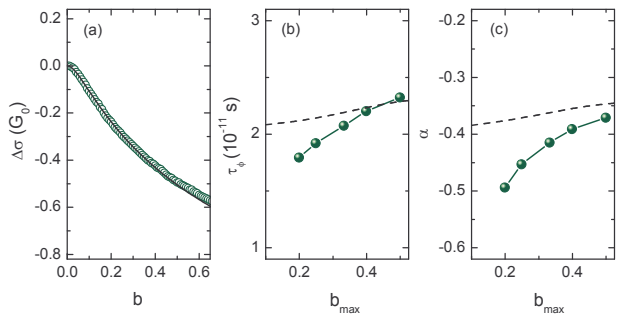}
\caption{(Color online)  (a) -- The magnetic field dependence of $\Delta\sigma$  for
$\sigma=46.0\,G_0$ measured at $T=1.35$~K. Symbols are the experimental data,
the curves are the best fit to Eq.~(\ref{eq30})  made within
the magnetic filed range $b=(0-b_\text{max})$, $b_\text{max}=0.3$.
(b) and (c) -- The dependence of the fitting parameters $\tau_\phi$ and $\alpha$, respectively,
on the upper limit of the fitting magnetic field range, $b_\text{max}$. The dashed curves in
(b) and (c) are the fitting parameters plotted against $b_\text{max}$
as they are obtained when  Eq.~(\ref{eq30}) is used for the fitting of the MC curve
calculated  for $\tau_\phi=2\times 10^{-11}$~s and
$\tau_\phi/\tau=30$ in the framework of the model valid beyond the diffusion regime
(for more details, see Ref.~\onlinecite{Note2}).
}\label{f3}
\end{figure}

Let us now consider the fitting results. In Fig.~\ref{f3}(a), we
present as an example the data measured for $\sigma=46.0\,G_0$ together
with the fitting curve. The parameters $\alpha$ and $\tau/\tau_\phi=x$
in Eq.~(\ref{eq30}) have been used as the fitting ones. The dephasing
time has been estimated with the use of the total conductivity and hole
density, $\sigma$ and $p$, respectively,  and average effective mass
$m_\text{av}$ as follows: $\tau_\phi=\tau/x$, where $\tau=\sigma
m_\text{av}/(e^2p)$, $m_\text{av}=(m_1+m_2)/2$ [see Fig.~\ref{f1}(d)].
The fitting magnetic field range is $b=(0-0.3)$. It is evident that
Eq.~(\ref{eq30}) describes  the MC curve rather well.

It should be mentioned that  the values of the fitting parameters are
somewhat sensitive to the width of the fitting $b$-interval as
Figures~\ref{f3}(b) and \ref{f3}(c) illustrate. The dephasing time
$\tau_\phi$ increases while the prefactor  $\alpha$ slightly decreases
in magnitude with the expanding  interval. This can be partially
attributed to the fact that the diffusion regime, $\tau\ll\tau_\phi$,
is not strictly satisfied under our experimental conditions:
$\tau_\phi$ is only $10-30$ times larger than $\tau$ depending upon the
conductivity and temperature. As shown in Ref.~\onlinecite{Note2} the
values of the fitting parameters are really dependent on the fitting
interval if the diffusion formula, Eq.~(\ref{eq30}), is used for
description of the MC curve beyond the diffusion regime. From the
dashed curves in Figs.~\ref{f3}(b) and \ref{f3}(c) it is evident that
these dependences are quantitatively close to that observed
experimentally.

Thus, the fitting value of the prefactor is close to $-0.5$ that
corresponds to the case when transition time between the spin split
subbands $\tau_{12}$ is less than $\tau_\phi$.

\begin{figure}
\includegraphics[width=0.9\linewidth,clip=true]{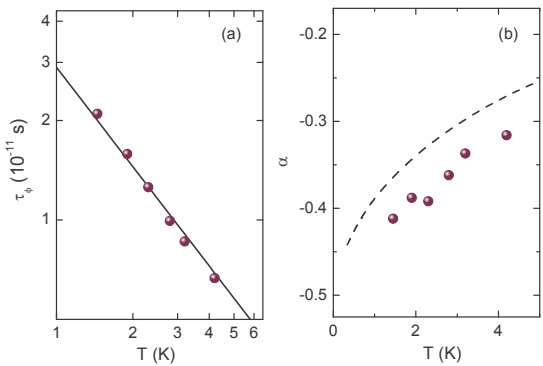}
\caption{(Color online) The temperature dependences of the fitting parameters $\tau_\phi$ (a) and
$\alpha$ (b) found in the fitting interval  $b=0-0.3$  for $p=7.9\times 10^{10}$~cm$^{-2}$,
$\sigma=63\,G_0$. Straight line in (a) is the dependence $\tau_\phi=2.9\times 10^{-11}/T$, s. The dashed curve in
(b) is $\alpha$ plotted against $T$ as it is obtained when the diffusion
expression Eq.~(\ref{eq30}) is used within the range $b=(0-0.3)$ for the fit of the MC curve
calculated beyond the diffusion approximation.\cite{Note2}
}\label{f4}
\end{figure}

Now, let us inspect Fig.~\ref{f4}, in which the temperature dependences
of $\tau_\phi$ and $\alpha$  are presented. It is seen that $\tau_\phi$
varies as $1/T$, as it should be when the inelasticity of \emph{e-e}
interaction is responsible for dephasing. But the prefactor $\alpha$
changes with the temperature also. Such temperature dependence of
$\alpha$ is described quite well when one takes into account the
decrease of $\tau_\phi/\tau$ with the temperature increase [see dashed
curve in Fig.~\ref{f4}(b)], which worsens applicability of the
diffusion approximation.\cite{Note2}

Thus, a sufficiently good agreement of the theoretical MC curve with
experimental one, the conventional behavior of the fitting parameter
$\tau_\phi$ with the temperature and understandable behavior of
$\alpha$, all this together testifies  the adequacy of the used model
to find the value of $\tau_\phi$. Such data treatment carried out
within wide hole density range shows analogous coincidence.

Let us now present the conductivity dependence of $\tau_\phi$. The
theory\cite{Nar02} predicts that the value of $\tau_\phi$ should
increase with the conductivity as $\sigma/\ln{\sigma}$, when the
inelasticity of \emph{e-e} interaction is the main mechanism of the
dephasing. Such a prediction is justified in experiments on the quantum
wells with ordinary spectrum (see, e.g., Ref.~\onlinecite{Min04-2},
where the data for GaAs/In$_{0.2}$Ga$_{0.8}$As/GaAs quantum well are
presented).

\begin{figure}
\includegraphics[width=1.0\linewidth,clip=true]{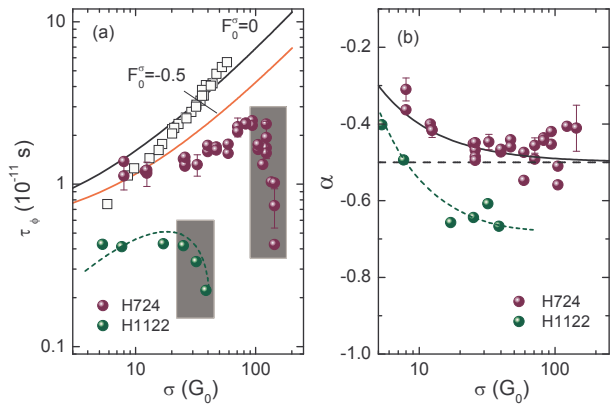}
\caption{(Color online) The conductivity dependence of $\tau_\phi$ (a) and $\alpha$
(b) for two heterostructures under study (spheres).
The squares are the data from Ref.~\onlinecite{Minkov13-1} relating to the electron gas in
HgTe-QW with $d=5$~nm. The solid curves in the panel (a) are calculated according to
Ref.~\onlinecite{Nar02}. The shadow areas indicate the drops in the $\tau_\phi$~vs~$\sigma$ dependences.
The solid curve in the panel (b) is the dependence
$-0.5\times(1-2\,G_0/\sigma)$.\cite{Min04-2}  $T=1.35$~K.  The dotted lines are provided as
a guide to the eye.
}\label{f5}
\end{figure}

The experimental  dependence $\tau_\phi(\sigma)$ measured at $T=1.35$~K
is shown in Fig.~\ref{f5}(a). Firstly, we consider the region where the
conductivity is less than $\simeq 100\,G_0$ [this corresponds to the
case when the distance between the valence band top and Fermi level is
less than $(20-25)$~meV]. It is evident that $\tau_\phi$ increases with
the conductivity within this conductivity range.  Such a behavior
agrees rather well with the theoretical prediction.\cite{Nar02}  The
absolute values of $\tau_\phi$ are also in satisfactory agreement with
the theoretical results obtained in Ref.~\onlinecite{Nar02}  that is
clearly seen from Fig.~\ref{f5}, where the solid curves represents the
calculation results for two values of the parameter of \emph{e-e}
interaction $F_0^\sigma$: $F_0^\sigma$=0 and $F_0^\sigma=-0.5$.

Note that the second fitting parameter $\alpha$ decreases in absolute
value with the decreasing conductivity [Fig.~\ref{f5}(b)]. Such a
behavior is also  natural and is in reasonable agreement with the
behavior of $\alpha$ in conventional two-dimensional systems and with
the theoretical dependence obtained with taken into account second loop
corrections:
$\alpha(\sigma)=\alpha(\sigma\to\infty)(1-2\,G_0/\sigma)$,\cite{Min04-2}
where $\alpha(\sigma\to\infty)=-1/2$ for our case.

It is appropriate at this point to recall again that the conductivity
dependences of $\tau_\phi$ found by the analogous manner for electrons
were different in the structures with inverted ($d>d_c$) and normal
($d<d_c$) spectra.\cite{Minkov12,Minkov13-1} It was found in
Ref.~\onlinecite{Minkov13-1} that the dephasing time of electrons
increased with growing $\sigma$ at $d<d_c$. Figure~\ref{f5}(a) shows
that in the structures investigated in the present paper [with
$d=(5.6-5.8)$~nm$<d_c$] the dephasing time of the holes behaves the
same. This is different from that observed for the electrons in the
structures with the inverted spectrum, where $\tau_\phi$ was
practically independent of $\sigma$ for whatever reason, which is as
yet unknown.\cite{Minkov12} The interference correction for holes with
the inverted spectrum still remains to be studied.

Thus, the  interference quantum correction to the conductivity in HgTe
quantum well with normal spectrum, $d<d_c$, both for electrons and for
holes, is analogous to that in ordinary 2D systems. Namely, the
magnetoconductivity curves are well described by the conventional
expression Eq.~(\ref{eq30}). The temperature and conductivity
dependences of $\tau_\phi$ found from the fit to Eq.~(\ref{eq30}) are
more or less close to the theoretical ones derived for the case when
inelasticity of \emph{e-e} interaction is the main dephasing mechanism.

Let us now consider the dephasing time at higher conductivity
($\sigma>100\,G_0$) that  corresponds to the hole density larger than
$1\times 10^{11}$~cm$^{-2}$ for the structure H724. It is seen from
Fig.~\ref{f5}(a) that the value of $\tau_\phi$ decreases abruptly with
the conductivity increase. As Fig.~\ref{f1}(a) shows the  hole density
$1\times 10^{11}$~cm$^{-2}$ is close to though somewhat less than the
value at which the increase of the hole density begins to slow down
with the increasing negative gate voltage. This suggests that both
these facts are of common origin. The analogous behavior of the
conductivity dependence of $\tau_\phi$ is observed in the structure
H1122 with narrower quantum well $d=5.6$~nm (see Fig.~\ref{f5}), in
which the change of slope in the $p_H$~vs~$V_g$ dependence  occurs at
higher hole density, $p\simeq 1.1\times 10^{11}$~cm$^{-2}$. Since the
change in $dp_\text{H}/dV_g$ at low $V_g$ values is possibly caused by
the population of the secondary maxima in the valence band spectrum
[see Fig.~\ref{f1}(c)], it is reasonable to assume that the appearance
of the  carriers in the secondary maxima leads to additional mechanism
of the dephasing due to inelasticity of \emph{e-e} interaction of the
holes in main maximum with these carriers.

Another possibility to explain the feature under discussion is
existence of localized states in the lower barrier which start to be
occupied with the decreasing gate voltage leading to the same effect in
the dependence $p(V_g)$ at $p\simeq 1\times 10^{11}$~cm$^{-2}$.
Inelasticity of the interaction with carriers in these states may also
result in the sharp decrease of $\tau_\phi$. We cannot exclude this
mechanism at the moment.

\section{Conclusion}

The results of experimental study of the interference quantum
correction to the conductivity in the narrow quantum well HgTe of the
hole type with the normal energy spectrum are presented. Analysis of
the interference induced low-field magnetoconductivity has been
performed with taking into account the strong spin-orbit splitting of
the hole subband. We have shown that the temperature dependence of the
phase relaxation time found from the fit of the magnetoconductivity
curves is close to $1/T$ over the whole conductivity range
$\sigma=(5-150)\,G_0$. Such a behavior  is typical for the dirty
two-dimensional systems at low temperature when the inelasticity of
electron-electron interaction is the main dephasing mechanism.

The conductivity dependence of the phase relaxation times is
nonmonotonic that may be  consequence of nonmonotonic dispersion $E(k)$
of the main hole subband of spatial quantization. At relatively low
conductivity ($\sigma<100\,G_0$ for the QW of $5.8$~nm width), when the
Fermi level lies above the secondary maxima of the dispersion, the
dephasing time increases with the conductivity increase analogously to
that observed for electrons in narrow HgTe quantum wells with the
normal energy spectrum\cite{Minkov13-1} and in ordinary single quantum
wells. Such a behavior is in agreement with that predicted
theoretically\cite{Nar02} for the case when inelasticity of \emph{e-e}
interaction is responsible for the phase relaxation. At the same time,
it differs markedly from the behavior of $\tau_\phi$ obtained in the
HgTe quantum wells with $d=(9-10)$~nm with the inverted energy
spectrum, where $\tau_\phi$ remains nearly constant over the wide
conductivity range.\cite{Minkov12} The $\tau_\phi$ decrease evident at
higher conductivity $\sigma>100\,G_0$, when the Fermi level is close or
even arrives the secondary maxima, may result from the additional
channel of the inelastic interaction between the carriers in the main
and secondary maxima.

\section*{Acknowledgments}

We are grateful to I.V.~Gornyi, V.Yu.~Kachorovskii, and P.M.~Ostrovsky
for useful discussions. The work has been supported in part by the RFBR
(Grant Nos. 12-02-00098 and 13-02-00322) and RAS (Project No.
12-P-2-1051). A.V.G. and O.E.R. gratefully acknowledge financial
support from the Ministry of education and science of Russia under
Project Nos. 3.571.204/K and 2014/236.

%\bibliography{QuantumCorrections}

%

\newpage

\section*{Supplemental Material}

\setcounter{figure}{0}

\setcounter{equation}{0}

\setcounter{table}{0}

\renewcommand{\theequation}{S.\arabic{equation}}
\renewcommand{\thefigure}{S.\arabic{figure}}
\renewcommand{\thetable}{S.\arabic{table}}

\subsection{Applicability of the one-type-carrier approximation to
analysis of weak antilocalization magnetoconductivity} \label{sec:sm1}

\subsubsection{Estimations of $\tau$ and $B_{tr}$ in subbands}
\label{ssec:sm1:1}

The analysis of the SdH oscillations  in the structures investigated
shows that the valence band is strongly split by spin-orbit interaction
so that the ratio of the hole densities in the subbands is
approximately equal to two. That is why the parameters determining the
weak localization  can be different as  well. As follows from the main
paper in order to obtain the phase relaxation time it is needed to
known the transport relaxation times $\tau_i$ and transport magnetic
fields $B_{tr}^{(i)}$, where $i=1,2$ numbers the spin split subband.

\begin{figure}
\includegraphics[width=0.6\linewidth,clip=true]{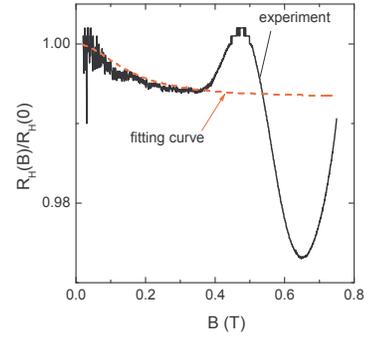}
\caption{(Color online) The magnetic field dependence of $R_\text{H}(B)/R_\text{H}(0)$ for
$\sigma=58.7\,G_0$. Structure H724. The solid curves are the data, the dashed ones are the results of the best fit
with the parameters given in Table~\ref{tab1sm}. }\label{f1sm}
\end{figure}

The $\tau_i$ and $B_{tr}^{(i)}$  values can be estimated from analysis
of the experimental magnetic field dependences of $\rho_{xx}$ and $R_H$
at classical magnetic field, $\mu_i B<1$, within framework of the
standard model of conductivity by two types of carriers. Because there
are additional mechanisms of the magnetic field dependence of
$\rho_{xx}$ (e.g., the quantum correction due to \emph{e-e}
interaction), we have analyzed only the dependence $R_H(B)$. It has
been fitted to the classical textbook expression for the Hall
coefficient\cite{Blatt} using mobilities $\mu_1$ and $\mu_2$ as the
fitting parameters, and $p_1$ and $p_2$ which were found from the SdH
oscillations [see Fig.~1(a) in the main paper]. As an example we
present the results of such a fit for $\sigma=58.7\,G_0$ in
Fig.~\ref{f1sm}. All the parameters used in and found from the fit are
listed in Table~\ref{tab1sm}. The transport relaxation times $\tau_i$
were found as $\tau_i=\mu_im_i/e$ with $m_1$ and $m_2$ from Fig.~1(d)
of the main paper. The transport magnetic fields has been calculated as
$B_{tr}^{(i)}=\hbar/(2e l_i^2)=\pi^3 \hbar p_i/(2e)\times
(G_0/\sigma_i)^2$.

Although as seen from Fig.~\ref{f1sm} the fit quality is quite good,
the accuracy in determination of the fitting parameters is not very
high. This is because the variation of the Hall coefficient in the
magnetic field is less than $1$~\% and the experimental $R_H$~vs~$B$
traces are noisy on this scale.  For this concrete case, we estimate
the error by the value $\pm 20\,\%$.

\begin{table}
\caption{The parameters of  the different spin subbands  for
$\sigma=58.7\,G_0$.}. \label{tab1sm}
\begin{ruledtabular}
\begin{tabular}{ccc}
 & $i=1$ & $i=2$ \\
 \colrule
 $p_i$ ($10^{10}$~cm$^{-2}$)\footnote{Found from the SdH oscillations}
  & 2.4 & $4.8$\\
 $\mu_i$ ($10^4$~cm$^{2}$/V~s)\footnote{Obtained from the fit of $R_{H}$~vs~$B$ data.} & $7.0$&  $5.9$\\
 $m_i/m_0$ &$0.012$& $0.028$\\
 $\tau_i$~($10^{-13}$~s) &$4.8$& $9.4$\\
 $B_{tr}^{(i)}$~(mT) &$5.1$& $3.6$\\
$B_{tr}$~(mT) &\multicolumn{2}{c}{$4.3$}\\
$\tau_\phi$~($10^{-11}$~s) &\multicolumn{2}{c}{$2.0$}\\
$\tau_\phi^\text{fit}$~($10^{-11}$~s) &\multicolumn{2}{c}{$2.17$}\\
$\alpha^\text{fit}$ &\multicolumn{2}{c}{$-0.98$}\\
 \end{tabular}
\end{ruledtabular}
\end{table}

\subsubsection{Estimation of errors at using one-band approximation
for analysis of WL MC} \label{ssec:sm1:2}

Let us estimate the error in determination of the $\tau_\phi$ and
$\alpha$ values, which  arises if one uses ``one-band'' expression
\begin{equation}
\Delta\sigma = \alpha G_0 {\cal H}\left( \frac{B}{B_{tr}}, \frac{\tau}{\tau_\phi}\right)
\label{eq10sm}
\end{equation}
to fit the data for the case of two strongly split spin subbands.

\begin{figure}
\includegraphics[width=0.58\linewidth,clip=true]{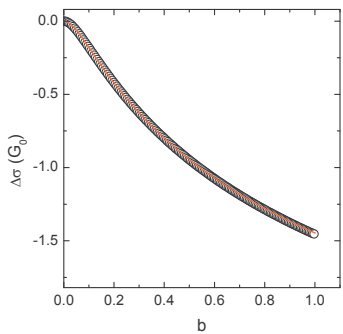}
\caption{(Color online) The magnetoconductivity plotted as a function of $b=B/B_{tr}$.
The symbols are calculated from Eq.~(2) of the main paper, the curves are results of the best fit by
Eq.~(\ref{eq10sm}) within the $b$ range from $0$ to $0.3$. The parameters are given in Table~\ref{tab1sm}.}\label{f2sm}
\end{figure}

We start with the case of slow transitions between spin subbands
($\tau_{12}\gg \tau_1, \tau_2$). We have calculated the interference
correction to the conductivity using Eq.~(2) from the main paper with
the parameters from Table~\ref{tab1sm}. Therewith, we have supposed
$\tau_\phi^{(1)}=\tau_\phi^{(2)}=\tau_\phi$ because  the dephasing time
is determined by conductivity under conditions that  inelasticity of
\emph{e-e} collisions is responsible for  dephasing. Then the
calculated $\Delta\sigma$~vs~$B$ curve  has been processed  by the same
way as experimental data, i.e., fitted to Eq.~(\ref{eq10sm}), in which
$B_{tr}=\pi^3 \hbar p/e\times (G_0/\sigma)^2$, $\tau=\sigma m/(e^2 p)$,
where $p$ stands for the total hole density $p_1+p_2$, $\sigma$ is the
total conductivity, and $m=(m_1+m_2)/2$. The parameters $\alpha$ and
$\tau_\phi$ have been used as  the fitting ones. It is evident from
Fig.~\ref{f2sm} that Eq.~(1) describes the simulated dependence
$\Delta\sigma(B)$ very well.

\begin{figure}
\includegraphics[width=0.95\linewidth,clip=true]{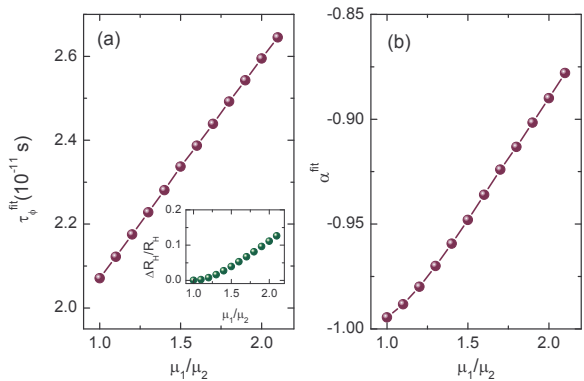}
\caption{(Color online) The fitting parameters $\tau_\phi^\text{fit}$ (a)
and $\alpha^\text{fit}$ (b) plotted against the
$\mu_1$ to $\mu_2$ ratio.}\label{f3sm}
\end{figure}

Since the accuracy in obtaining mobilities is not high under our
experimental conditions, it is useful to estimate how strongly the
fitting parameters $\tau_\phi^\text{fit}$ and $\alpha^\text{fit}$
depend on the $\mu_1$ to $\mu_2$ ratio. For this purpose we have
calculated the set of the magnetoconductivity curves $\Delta\sigma(B)$
with $\tau_\phi=2\times 10^{-10}$~s  for the different $\mu_1$ to
$\mu_2$ ratio values while keeping $n_1=2.4\times 10^{10}$~cm$^{-2}$,
$n_2=4.8\times 10^{10}$~cm$^{-2}$, and $\sigma=58.7\,G_0$. Note the
change of the Hall coefficient in the magnetic field $\Delta
R_H/R_H=[R_H(0)-R_H(\infty)]/R_H(0)$ does not exceed $15$~\% therewith
[see the inset in Fig.~\ref{f3sm}(a)]. Then, we have performed the
fitting procedure  within the $b$ range from $0$ to $0.3$. The
parameters $\tau_\phi^\text{fit}$ and $\alpha^\text{fit}$ corresponding
to the best fit are presented in Figs.~\ref{f3sm}(a) and \ref{f3sm}(b),
respectively. It is seen that the error in determination of $\tau_\phi$
does not exceed $30$~\%, while the $\mu_1$ to $\mu_2$ ratio varies
within the relatively wide interval: $\mu_1/\mu_2\simeq 1-2$. The value
of  $\alpha^\text{fit}$ remains always close to $-1$.

Let us now turn to the case of fast transitions between subbands
($\tau_{12}\ll \tau_1, \tau_2$). In this regime the relationship
between the fitting and used values of $\tau_\phi$ can be obtained
analytically because $\Delta\sigma(B)$ is given by Eq.~(\ref{eq10sm})
in which $\alpha$ is equal to $-1/2$ exactly and $\tau_\phi$ is
replaced by some effective value.\cite{Averkiev98SM} Our analysis shows
that the error in $\tau_\phi$  in this case does not exceed the value
of $10$~\%.

Thus, we conclude that the use of Eq.~(\ref{eq10sm}) for description of
the interference induced magnetoconductivity is justified under our
experimental conditions. Therewith, the fitting procedure allows us to
obtain not only the value of the phase relaxation time  but to estimate
the rate of  intersubband transitions. If the value of
$\alpha^\text{fit}$ is close to $-1/2$, so $\tau_{12}\ll \tau_{\phi}$.
When $\alpha^\text{fit}\simeq -1$, $\tau_{12}\gg \tau_{\phi}$.

\subsection{Workability of the diffusion formula beyond diffusion
regime} \label{sec:sm2}

\begin{figure}[t]
\includegraphics[width=\linewidth,clip=true]{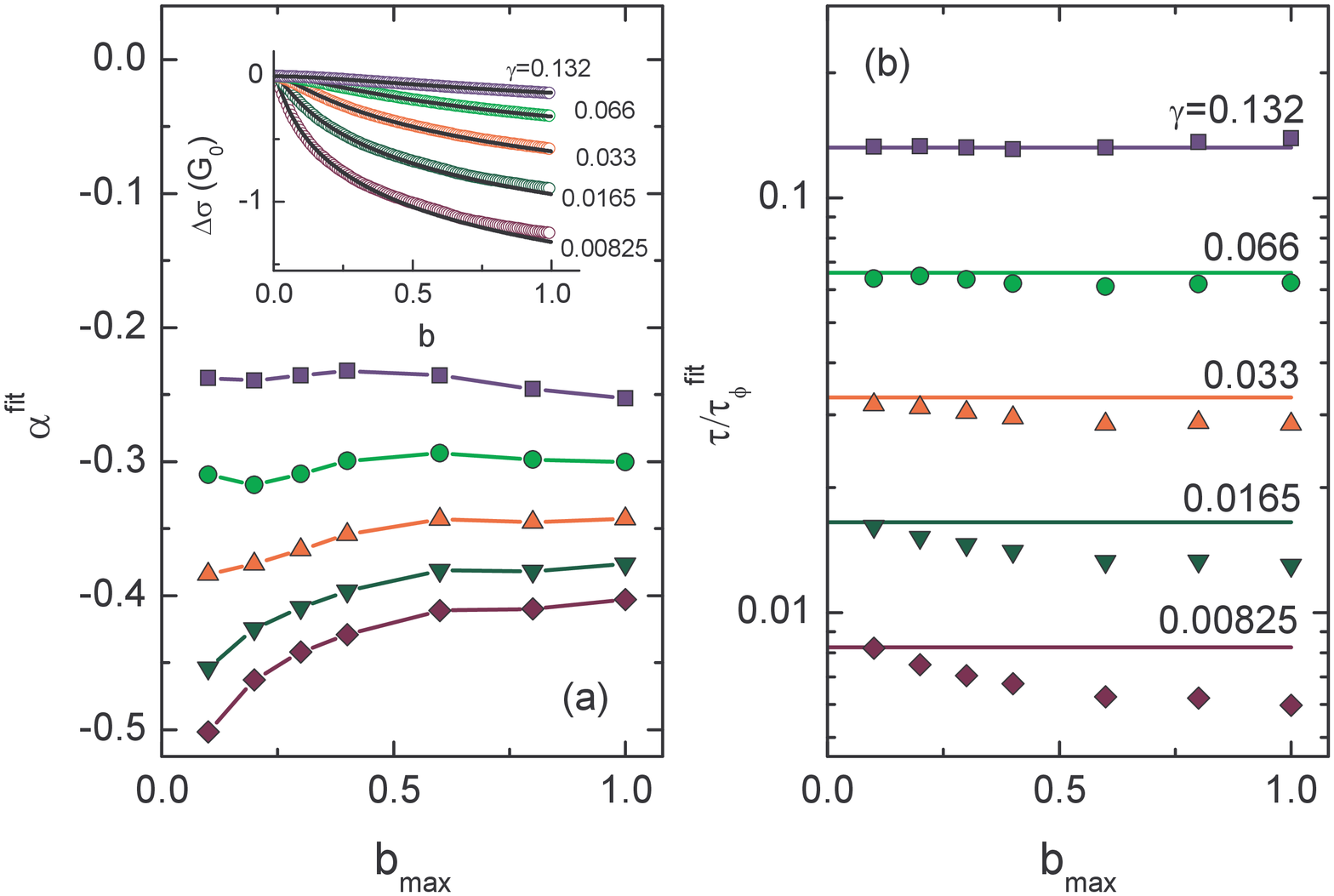}
\caption{(Color online)  The fitting parameters $\alpha^\text{fit}$ (a)
and $\tau/\tau_\phi^\text{fit}$ (b)  plotted against the width
of magnetic field range, in which the fit of the data presented in the inset in panel (a) has been done.
In the inset: symbols are obtained from the numerical simulation with
different values of $\gamma=\tau/\tau_\phi$; the curves are the results of the best fit by Eq.~(\ref{eq10sm})
within the range $b=(0-b_\text{max})$, $b_\text{max}=0.3$.}\label{f4sm}
\end{figure}

Equations (2) and (3) from the main paper, as well Eq.~(\ref{eq10sm}),
are valid in the diffusion regime, i.e., when both of the two
conditions $B\ll B_{tr}$ and $\tau_\phi\gg \tau$ are satisfied. The
calculation of WL MC beyond the diffusion regime was carried out in a
number of papers (e.g., in
Refs.~\onlinecite{Wit87,Dmit97,Cas94,Golub05,Porubaev13}). However the
expressions obtained  are so cumbersome that it is difficult to use
them for the fit of experimental curves. In order to estimate how well
is the fitting parameters obtained  when Eq.~(\ref{eq10sm}) is used for
the description  of the MC curve if the conditions of the diffusion
regime are satisfied not strictly,  we have used the numerical
simulation approach developed in Ref.~\onlinecite{Min00-1SM}. The
``experimental'' curves have been calculated  with the use  of Eq.~(21)
from that paper. In order to take into account the fast transitions
between spin subbands, $\tau_{12}\ll\tau_\phi$, the right-hand-side of
this equation has been multiplied by the factor $-1/2$ (for more
details, see Ref.~\onlinecite{Germ07-1}). The simulated data for
different values of $\gamma=\tau/\tau_\phi$ and fitting curves  are
presented in the inset in Fig.~\ref{f4sm}(a). It is evident that the
data are fitted by Eq.~(3) perfectly. As Figures~\ref{f4sm}(a) and
\ref{f4sm}(b) illustrate the values of the fitting parameters $\alpha$
and $\tau/\tau_\phi$ are sensitive to the width of magnetic field range
$(0-b_\text{max})$, in which the fit is performed. However, inspection
of Fig.~\ref{f4sm}(b) shows that the relative difference between the
$\tau/\tau_\phi^\text{fit}$ values and the values $\tau/\tau_\phi$ used
in the simulation procedure does not exceed $10\,\%$, if one restricts
the fitting interval by the value $b_\text{max}=0.3$.

\end{document}